\documentclass[letterpaper,11pt,twoside]{report}

\pdfoutput=1

\title
{
	On Implementation of a Safer C Library, ISO/IEC TR 24731.\\
	Technical Report\\
}

{\author
	{\bf
		{CIISE Security Investigation Initiative}\\\hline\\
		Represented by:\\\\
		Marc-Andr\'e Laverdi\`ere-Papineau\\
		Serguei A. Mokhov\\
		Djamel Benredjem\\
		\texttt{\{ma\_laver,mokhov,d\_benred\}@ciise.concordia.ca}
		\\\\\\
		Montr\'eal, Qu\'ebec, Canada\\\\\\
	}
}

\date{April 2006}

\usepackage{graphicx}
\usepackage{latexsym}
\usepackage{makeidx}
\usepackage{url}
\usepackage{listings}
\usepackage{hyperref}

\makeindex

\newcommand{\xf}[1]{Figure~\ref{#1}}

\newcommand{\xs}[1]{Section~\ref{#1}}

\newcommand{\xl}[1]{Listing~\ref{#1}}

\newcommand{\file}[1]{\texttt{#1}\index{Files!#1}}
\newcommand{\tool}[1]{\texttt{#1}\index{Tools!#1}}

\newcommand{\api}[1]{\texttt{#1}\index{API!#1}}

\newcommand{\lucidL}[1]{{$\mathit{Lucid}$}($L$) }

		{}

\def\myvert{\raise 2.27pt \hbox{\vrule depth 0pt height 8pt width 0.2mm}}
\def\myarrow{\hspace*{0.43mm}%
             \raise 2.29pt\hbox{\vrule depth 0pt height 8pt width 0.16mm}%
             \hspace*{-0.32mm}%
             $\longrightarrow$
             \ %
             }

\lstdefinestyle{patchStyle}{
    captionpos=b,%
    showstringspaces=false,%
    showspaces=false,
    frame=single,  
    extendedchars=true,%
    basicstyle=\scriptsize\tt, 
    linewidth=1\linewidth,%
    language=[ANSI]C,%
    breaklines=true,
    float=phtb,  
}

\lstdefinestyle{synopsysStyle}{
    captionpos=t,%
    showstringspaces=false,%
    showspaces=false,
    frame=single,  
    framerule=2pt,
    extendedchars=true,%
    basicstyle=\small\tt, 
    linewidth=1\linewidth,%
    language=[ANSI]C,%
    breaklines=true,
    float=phtb,  
}

\begin{document}

	\pagestyle{empty}

	\begin{titlepage}
		\maketitle
	\end{titlepage}

	\cleardoublepage

	\pagestyle{myheadings}
	\markboth {\hfill On Implementation of a Safer C Library, ISO/IEC TR 24731.}
					{On Implementation of a Safer C Library, ISO/IEC TR 24731.}

	\pagenumbering{roman}
\tableofcontents
\clearpage
\pagenumbering{arabic}


	\chapter{Introduction}
\index{Introduction}
\label{chapt:intro}

\section{Security Problems in C Standard Functions}\label{sect:intro:problems}

The functions standardized as part of ISO C 1999 and their addendums
improved very little the security options from the previously available library.

The largest flaw remained that no function asked for the buffer size of destination buffers
for any function copying data into a user-supplied buffer. According to earlier research
we performed \cite{ourselves}, we know that error condition handling was the first solution to
security vulnerabilities, followed by precondition validation. The standard C functions typically
perform little precondition validation and error handling, allowing for a wide range of security
issues to be introduced in their use.

For example:\\
\small
\begin{tabular}{|p{4.5in}|p{2in}|}
\hline
\texttt{char *strncat(char *dest, const char *src, size\_t n);} & does not null-terminate, can still overflow  \\ \hline
\texttt{char *strtok(char * restrict s1, const char * restrict s2);} & not reentrant\\ \hline
\texttt{size\_t strlen(const char *s);} & can iterate in the memory up to an invalid page and cause a program crash \\\hline
\end{tabular}
\normalsize

This effort remained not enough, and many projects developed additional functions, namely:
	\begin{itemize}
	\item OpenBSD \api{strlcpy} family \cite{strlcpy}
	\item GNU C extensions \cite{gnuext}
	\item Microsoft \file{strsafe.h} and others \cite{strsafe}
	\end{itemize}

\section{Introducing ISO/IEC}\label{sect:intro:iso}

The International Standardization Organization (ISO) and the
International Electrotechnical Commission (IEC) are
standard-making bodies headquartered in Geneva (Switzerland) \cite{isointro}.

Both organizations are constituted from an international membership, with
local member organizations involved in standard-making activities as well. For example,
we have the Standards Council of Canada, American National Standards Institute,
Deutsches Institut f\"ur Normung, and Association fran\c{c}aise de normalisation \cite{isomembers}.

ISO and IEC collaborate closely on standards related to computer equipment and information technologies.

These organizations established a hierarchical structure under JTC 1 - Joint Technical Committee on Information Technology.
JTC 1 is subdivided in 17 subcommittees, one of which (SC 22) deals with programming languages, with
a working group for each programming languages \cite{jtc1}.

The C language is normalized by ISO/IEC JTC 1/ SC 22/ WG 14.
Its members include representatives from Microsoft, SEI/CMU, Cisco, Intel, etc \cite{wg14minutes}.
The Computer Security Laboratory of CIISE, through Pr. Debbabi, is a member of this Working Group
as Canadian representative with voting rights.

\section{ISO/IEC TR 24731}\label{sect:intro:tr}

In the ISO jargon, TR 24731 \cite{tr} is a Technical Report Type 2 \cite{isotechguidelines, isotrtypes}, which means that the
document is not a standard, but a direction for future normalization. This specification is currently
in the draft state.

Titled ``TR 24731: Safer C library functions'', it defines 41 new library functions for
memory copying, string handling (both for normal and wide character strings), time printing,
sorting, searching etc. Another inovation it brings is a constraint handling architecture, forcing error handling when
certain security-related preconditions are violated when the functions are called. It also
specifies the null-termination of all strings manipulated through its function and introduces
a new unsigned integer type that helps preventing integer overflows and underflows. It is currently implemented by Microsoft as part of their Visual Studio 2005 \cite{seacord05securecoding}.


	\chapter{Architecture}
\index{Architecture}


In this chapter, we examine the architecture of our implementation of ISO/IEC TR 24731. We first introduce our 
architectural philosophy before informing 
the reader about the Siemens Four View Model, an architectural methodology for the conception of large-scale software systems.

Afterwards, we examine each of the view, as architected for our library.

Finally, we conclude with other software engineering matters that were of high importance
in the development of our implementation.

\section{Principles and Philosophy}\label{sect:architecture:principles}
\index{Methodology!Principles and Philosophy}

The library specification imposes that the functions be in addition of other standard
functions, in the same header files.
However, we do not want our implementation to re-implement the standard C library, nor
do we want to augment an existing implementation and be bound to a specific platform.
The compromise solution is to organize the code to be using
the low-level implementation of any existing C library, such as the one from GNU \cite{glibc}, FreeBSD or OpenBSD.

In short, our principles are:
\begin{description}
\item [Platform independence] We target systems complying with the POSIX standard.
\item [Standards Compliance] We will implement the library using features available only in ISO C99 and POSIX.
\item [C Library Independence] We will architect in a way that prevents us being tied to the underlying C library.
\item [Realistic Compiler Indepedence] A corollary of standards compliance, we will avoid compiler-specific macros and optimizations as possible. This means that the source code should be free of such dependencies, but that the build
process may be bound to the compiler.
\item [Reasonable Efficiency] We will architect and implement an efficient library, but will avoid advanced programming
tricks that improve the efficiency at the cost of maintainability and readability.
\item [Simplicity And Maintainability]  We will target a simplistic and easy to maintain organization of the source.
\item [Architectural Consistency] We will consistently implement our architectural approach. We could say that we will have a ``template'' approach.
\item [Separation of Concerns] We will isolate separate concerns between modules and within modules to encourage reuse and code simplicity.
\item [Functional Grouping] We will have functional coupling between within a module, meaning that functions will
have a commonality of type, such as I/O, string manipulations, etc.
\end{description}

\section{Summary of Siemens Four View Model}
\label{sect:architecture:siemens}
\index{Siemens Four View Model}

We decided to use the Siemens Four View Model for our architectural description, mostly
due to previous experience using the technology. A detailed description with case studies
can be found in \cite{asa00}. This methodology is introduced by scientists of Siemens Corporate Research
and has been successfully applied to a variety of systems, many of which with real-time and embedded requirements.
Please refer to \xf{fig:architecture:chalin} for an abstract presentation of the view.
In the context of this project report, we inform the reader of the basic principles of each view.

\begin{figure}[thb]
\includegraphics[height=4.5in, keepaspectratio=true]{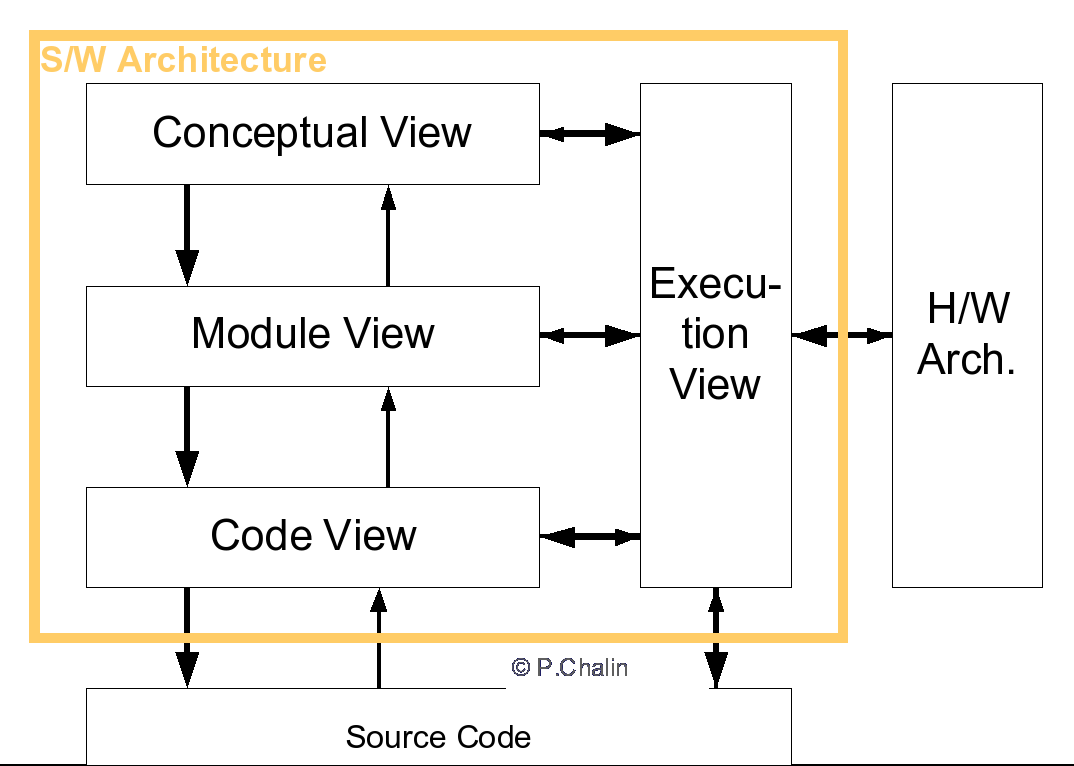}
\caption{High-Level Description of the Four Views \cite{chalin}}
\label{fig:architecture:chalin}
\end{figure}

\subsection{Conceptual View}
\index{Siemens Four View Model!Conceptual View}

The conceptual view defines the conceptual components and their conceptual connectors, as
well as their conceptual configuration. The architect, in this phase, must also specify
resource budgets.

\subsection{Module View}
\index{Siemens Four View Model!Module View}

In the module view, the architect maps the conceptually-defined elements into modules and layers.
The architect must then define the interface to the modules.

\subsection{Execution View}
\index{Siemens Four View Model!Execution View}

In the execution view, the architect must define the runtime entities, the communication paths
and the execution configuration. It essentially means the mapping of modules to threads and
processes, and to define the inter-process communication mechanisms to be used.
In the case of our library, there are no threads of execution per se,
as such, this architectural step was skipped.

\subsection{Code View}
\index{Siemens Four View Model!Code View}

In the code view, the architect must define the source components, intermediate components
and deployment components, followed by the build procedure and configuration management.

\subsection{Conceptual View}
\subsubsection{Conceptual Overview}
We divided our architecture in a few modules, isolating the core functionalities
from each library and grouping functions per library. We also decided to use wrappers
to a C library implementation. This overview can be seen in Figure \xf{fig:architecture:conceptual}.
\begin{figure}
\centering
\includegraphics[angle=-90, width=4in, keepaspectratio=true]{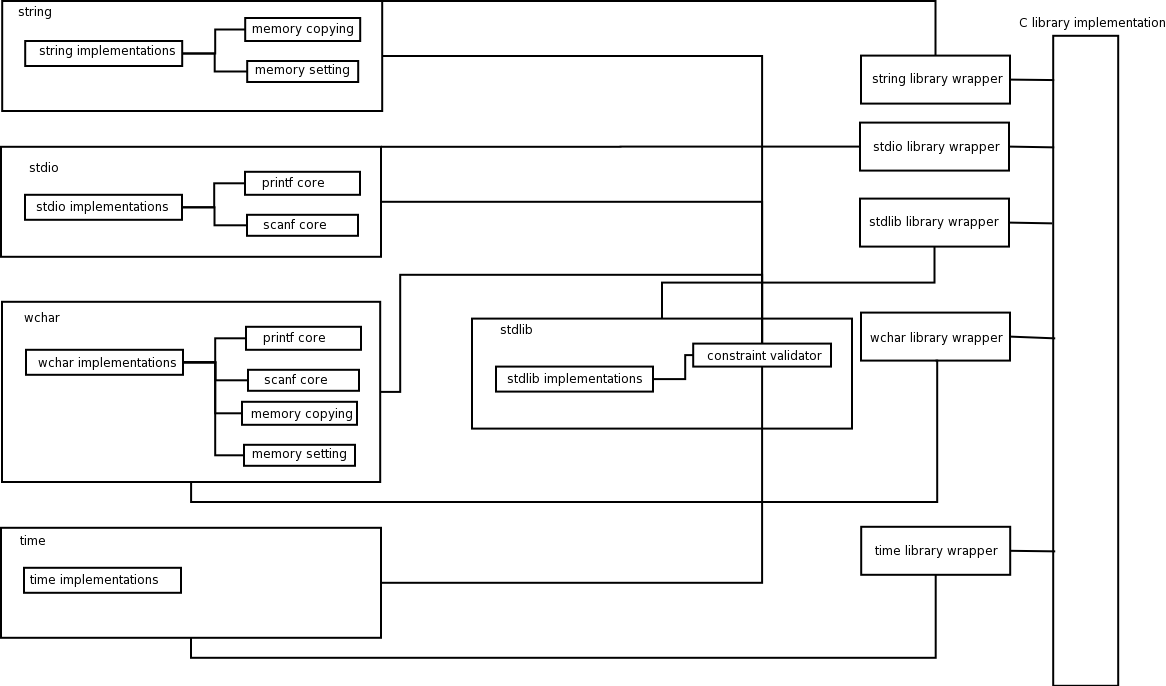}
\caption{Conceptual Modules}
\label{fig:architecture:conceptual}
\end{figure}

\subsubsection{Configurations}
The only options in configuration relate to the wrappers to use for a specific C library. Since this is decided at
compilation time and that only one is possible in any case, we decided not to further specify configurations.
\subsubsection{Protocols}
Because of the simplistic nature of the components (direction function calls) and that all
components need to be re-entrant, we conclude that no protocols are necessary.
\subsubsection{Resource Budgeting}
We did not specify any resource budget for any components due to the lack of specific constraints.

\subsection{Module View}
\subsubsection{Layering}
We divided our library between layers when we found significant redundancy for some operations. We decided to keep the
memory copying for wide characters apart from the one without wide characters due to the risk of integer overflows
that could result in faulty logic, and thus deserving a centralization of the functionality.

The complete view of our layering is included in Figure \xf{fig:architecture:module}.
\begin{figure}
\centering
\includegraphics[angle=-90, width=4in, keepaspectratio=true]{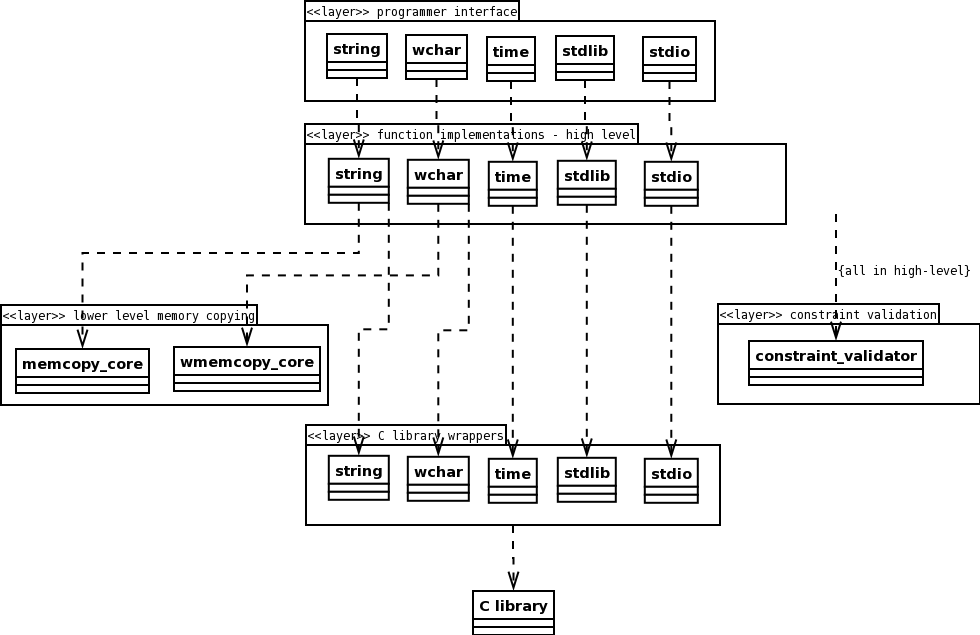}
\caption{Layering}
\label{fig:architecture:module}
\end{figure}

\subsubsection{Interface Design}

The only interface that needed to be designed was related to the constraint validation. Please refer to 
the specification given in section \xs{sect:constraints-api}

\section{Execution View}
\subsection{Runtime Entities}
In the case of our library, there are no threads of execution per se, as the thread is provided
by the calling program(s).
\subsection{Communication Paths}
It was resolved that the functions would all communicate through message passing.
\subsection{Execution Configuration}
Coherently with previous decisions, there are no specific execution configurations.

\section{Code View}
\label{sect:architecture:code}

\subsection{Source Components}

\begin{figure}
\centering
\includegraphics[height=4.5in, keepaspectratio=true]{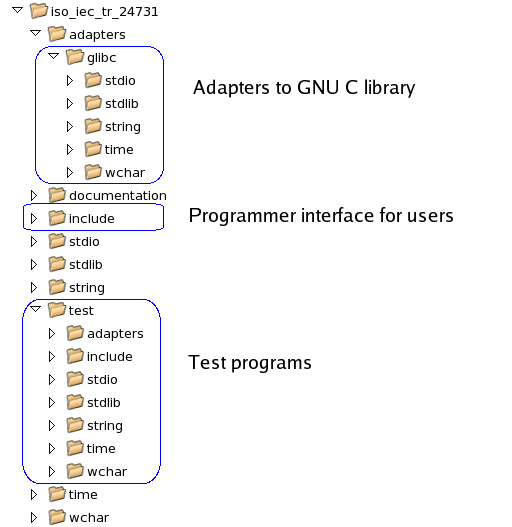}
\caption{Directory Tree}
\label{fig:architecture:tree}
\end{figure}

Each of \file{stdio.h}, \file{stdlib.h}, \file{string.h}, \file{time.h}, and \file{wchar.h} are mapped into a corresponding
directory, as a module. The folder named test copies the previous structure, and contains test programs exclusively.
The \file{include} directory contains the \file{.h} files to be included by external programs linking to
the library.  The folder named \file{adapters} holds a sub-directory per C library implementation to adapt to, and
each of those adapters mirror the base directory structure.
Each interface function was implemented in a file with its name, to facilitate maintainability.
Related functions are grouped in the file to which it is the most logically related.

We established a naming convention for functions as follows:
\begin{description}
\item[function] The function's interface for the user.
\item[\_\_function\_impl] The function's implementation, used within the library.
\item[\_\_function\_validate\_preconditions] The function's precondition validation component.
\item[\_\_function\_component] refactored sub-component of a function or common code within functions.
\end{description}

\subsection{Intermediate Components}

In order to facilitate the building process, we decided that each directory of functionality and adapters
will assemble all intermediary (\file{.o}) files into an archive (\file{.a}).

\subsection{Deployment Components}

The only deployment component will be a \file{.so} file that includes our implementation and the underlying library.

\subsection{Make Process}

The build process is organized as a hierarchical organization of makefiles. Each \file{makefile}
cross-reference the makefiles for its dependencies.
The intermediary objects defined are the standard \file{.o} that are also grouped in an archive (\file{.a}).
The testers are built separately from the library itself for efficiency and faster compilation.

The make files are designed to adapt to both Linux and Windows/Cygwin platforms by detecting the 
presence of Cygwin and using different compiler options consequently.

The documentation generation fits outside of the normal make process and is generated by the doxygen tool
itself.

\subsection{Configuration Management}

We used Subversion (\tool{svn}) \cite{svn} in order to manage the source code, makefile,
and documentation revisions.

\section{Example for One Module}

In \xf{fig:architecture:calldep} and \xf{fig:architecture:filedep}, we show the example of
function call and source file dependencies for a function implementing vfprintf\_s.

\label{sect:architecture:example}

\begin{figure}[htb]
\includegraphics[width=5in, keepaspectratio=true]{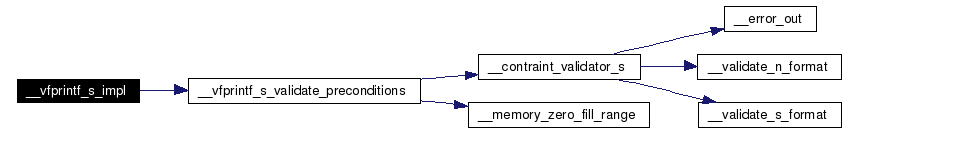}
\caption{Call dependencies for function vfprintf\_s}
\label{fig:architecture:calldep}
\end{figure}

\begin{figure}[htb]
\includegraphics[width=4in, keepaspectratio=true]{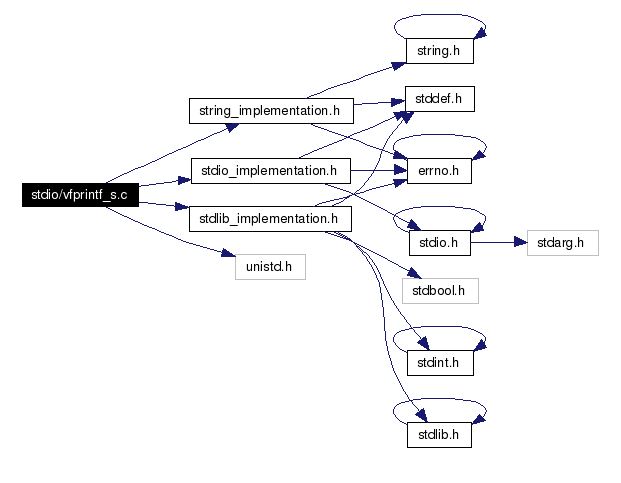}
\caption{File dependencies for function vfprintf\_s}
\label{fig:architecture:filedep}
\end{figure}

\section{Iterations}
\label{sect:architecture:iterations}

In order to efficiently reach our architectural goal, we divided the 
final objectives of the project in the following steps:

\begin{enumerate}
\item Implementation of the body, without precondition validations
\item Implementation of precondition validation and integration
\item Validation of compliance for all testable cases
\item Implementation of adapters to a C library
\item Redefinition of insecure function calls to our function calls
\end{enumerate}

\section{Coding Standards}
\label{sect:standards}

In order to produce high-quality code, we decided to normalize on the OpenBSD style.
We also decided to use Doxygen \cite{doxygen} source code documentation style for
its completeness and the automated tool support.


	\chapter{Implementation}
\index{Implementation}

This chapter gives concrete implementation details
for the constraint handling API and examples of its
usage.
An excessive amount of work was done to have a decent
precondition validation and this chapter mostly focuses
on the API aspect of this implementatition. A particular
achievement was to fully enable parsing and restraining
of \%s and \%n modifiers with the flex-based scanner.

%
%

\section{Run-time Constraint Handling API}
\label{sect:constraints-api}

For precondition validation we deviced an API to encasulate
run-time constraint check and violation information and an appropriate currently
registered constraint handler. Later on this was abstracted
with inline function calls reducing the clutter for the libc\_s
programmers such as ourselves and for those who might
maintain it after us.

The standard defines this type to allow custom constraint
handlers in \file{stdlib.h}:

\begin{lstlisting}[
    label={list:handling-api},
    caption={Standard API for Constraint Handling},%
    style=patchStyle
    ]
typedef void (*constraint_handler_t)(const char* restrict msg, void* restrict ptr, errno_t error);
constraint_handler_t set_constraint_handler_s(constraint_handler_t handler);
void abort_handler_s(const char * restrict msg, void * restrict ptr, errno_t error);
void ignore_handler_s(const char * restrict msg, void * restrict ptr, errno_t error);
\end{lstlisting}

In our implementation we mark \api{abort\_handler\_s} as the
default handler for all contraint violations. That means, after
erroring out, the handler calls \api{exit(0)} and the application
build around \tool{libc\_s} terminates.

\section{Constraint Violation Information Encapsulation API}
\label{sect:error-api}

This is the capsule enclosing the error information we defined.
The structure depicts some meta information about a paramater
and its value. An instance of this structure is created for
each input parameter to be validated.

\begin{lstlisting}[
    label={list:param-validation-status},
    caption={Synopsys: \api{param\_validation\_status\_t}},%
    style=synopsysStyle
    ]
#include "stdlib_implementation.h"

typedef struct _param_validation_status_t
{
	e_errcheck errtype;
	const void* value;
	const void* pairvalue;
	const void* result;
	const char * restrict function_name;
	const char * restrict param_name;
	const char * restrict pair_param_name;
	bool error_present;
} param_validation_status_t;
\end{lstlisting}

Some helper data structures allow us to describe more complex types.
The \api{param\_range\_t} struct allows specifying the range for
a domain value. A reference to the instance of this struct is passed
in the \api{param\_validation\_status\_t.pairvalue} field.
The \api{object\_range\_t} struct is supposed to contain the data
describing a memory objec with its starting address and length.
The purpose of this is to help with validation of ranged objects
that they do not overlap in memory. The implementor of the library
should provide the two intances of this struct as references
in \api{value} and \api{pairvalue} of the two ranged objects.
Which object reference goes to where is unimportant as the implementation
takes care of figuring out the object precedence.

\begin{lstlisting}[
    label={list:param-range},
    caption={Synopsys: \api{param\_range\_t}},%
    style=synopsysStyle
    ]
typedef struct _param_range_t
{
	size_t min;
	size_t max;
} param_range_t;
\end{lstlisting}

\begin{lstlisting}[
    label={list:object-range},
    caption={Synopsys: \api{object\_range\_t}},%
    style=synopsysStyle
    ]
typedef struct _object_range_t
{
	const void* restrict object_ptr;
	size_t object_length;
} object_range_t;
\end{lstlisting}

\section{Constraint Enumeration and Validator}
\label{sect:validator-api}

The enumeration in \xl{list:errcheck} defines most common error types
to check for and report. These correspond to the index for the human
readable error messages.

\begin{lstlisting}[
    label={list:errcheck},
    caption={Enumeration of most common error types to check for.},%
    style=patchStyle
    ]
typedef enum
{
	E_NOERROR = 0,
	E_NULL_PARAMETER_NOT_ALLOWED = 1,
	E_PARAMETER_OUT_OF_RANGE = 2,
	E_ENVIRONMENTAL_LIMIT_NOT_MET = 3,
	E_INVALID_FORMAT_PARAMETER_S = 4, /* %s */
	E_INVALID_FORMAT_PARAMETER_N = 5, /* %n */
	E_RSIZE_MAX_EXCEEDED = 6,
	E_NOT_ZERO = 7,
	E_OBJECTS_OVERLAP = 8,
	E_NOT_IMPLEMENTED = 9,
	E_TOKEN_END_NOT_FOUND = 10
} e_errcheck;
\end{lstlisting}

\begin{lstlisting}[
    label={list:validator},
    caption={Constraint Validator API},%
    style=patchStyle
    ]
errno_t __error_out(param_validation_status_t * restrict status);
errno_t __contraint_validator_s(param_validation_status_t * restrict status);
errno_t
__constraint_validator_object_overlap(const char * restrict function_name, const char * restrict parameterNames, const void * restrict object1Start, const size_t object1Size, const void * restrict object2Start, const size_t object2Size);
errno_t
__constraint_validator_value_inrange(const char * restrict function_name, const char * restrict parameterName, const size_t value, const size_t lowerBound, const size_t upperBound);
errno_t
__constraint_validator_not_null(const char * restrict function_name, const char * restrict parameter_name, const void * restrict value_ptr);
errno_t
__constraint_validator_not_null_args(const char * restrict function_name, const char * restrict parameter_name, const char * restrict format, const va_list args);
errno_t
__constraint_validator_s_format(const char * restrict function_name, const char * restrict parameter_name, const char * restrict format, const va_list args);
errno_t
__constraint_validator_n_format(const char * restrict function_name, const char * restrict parameter_name, const char * restrict format, const va_list args);
errno_t
__constraint_validator_not_zero(const char * restrict function_name, const char * restrict parameter_name, const size_t value);
errno_t
__constraint_validator_rsize_limit(const char * restrict function_name, const char * restrict parameter_name, const rsize_t value);
void
__report_constraint_violation_end_of_token_not_present(const char * restrict function_name, const char * restrict parameter_name);
\end{lstlisting}

\clearpage

\section{Constraint Handling Example}
\label{sect:example-1}

Our implementation of the API (in \xl{list:validator}) does something 
similar to the code snippet presented in \xl{list:query-for-check}.

\begin{lstlisting}[
    label={list:query-for-check},
    caption={Validation Code},%
    style=patchStyle
    ]
    ...
	param_validation_status_t format_string_validation;
	__memory_zero_fill_range(&format_string_validation, sizeof(param_validation_status_t));
	format_string_validation.errtype = E_INVALID_FORMAT_PARAMETER_S;
	format_string_validation.value = format;
	format_string_validation.pairvalue = &arg;
	format_string_validation.param_name = "format/arg null %s";
	format_string_validation.function_name = "vfprintf_s";

	error = __contraint_validator_s(&format_string_validation);

	if(error != OK)
	{
		errno = error;
		return error;
	}
	...
\end{lstlisting}



	\chapter{Results}
\index{Results}

This chapter summarizes the result achieved as of this writing.
This includes implemented API to this point as well as some
concrete results demonstraiting correctness of implementation.

\section{Implemented API}
\index{Implemented API}

This is the summary of the implemented API from the library and
our internal constraint handling.
We summarized the functions and data types added in ISO/IEC TR 24731 in this section for the sake of
reference.

\subsection{Library}
\index{Implemented API!Library}

\subsubsection{Data Types}

Added the following data types:
\api{rsize\_t},
\api{errno\_t},
\api{constraint\_handler\_t} that were necessary to add.

\subsubsection{Functions}

The functions in \xl{list:impl-funcs} were to the large extend implemented
by our team as of this writing. Likewise, \xl{list:not-impl-funcs} lists
API not yet addressed. Finally, \xl{list:partial-impl-funcs} lists API
implemented half-way through.

\begin{lstlisting}[
    label={list:impl-funcs},
    caption={Implemented Safer C Library API},%
    style=patchStyle
    ]
int fprintf_s(FILE * restrict stream, const char * restrict format, ...);
int fscanf_s(FILE * restrict stream, const char * restrict format, ...);
int printf_s(const char * restrict format, ...);
int scanf_s(const char * restrict format, ...);
int snprintf_s(char * restrict s, rsize_t n, const char * restrict format, ...);
int sprintf_s(char * restrict s, rsize_t n, const char * restrict format, ...);
int sscanf_s(const char * restrict s, const char * restrict format, ...);
int vfprintf_s(FILE * restrict stream, const char * restrict format, va_list arg);
int vfscanf_s(FILE * restrict stream, const char * restrict format, va_list arg);
int vprintf_s(const char * restrict format, va_list arg);
int vscanf_s(const char * restrict format, va_list arg);
int vsnprintf_s(char * restrict s, rsize_t n, const char * restrict format, va_list arg);
int vsprintf_s(char * restrict s, rsize_t n, const char * restrict format, va_list arg);
int vsscanf_s(const char * restrict s, const char * restrict format, va_list arg);
constraint_handler_t set_constraint_handler_s(constraint_handler_t handler);
void abort_handler_s(const char * restrict msg, void * restrict ptr, errno_t error);
void ignore_handler_s(const char * restrict msg, void * restrict ptr, errno_t error);
errno_t wctomb_s(int * restrict status, char * restrict s, rsize_t smax, wchar_t wc);
errno_t mbstowcs_s(size_t * restrict retval, wchar_t * restrict dst, rsize_t dstmax, const char * restrict src, rsize_t len);
errno_t wcstombs_s(size_t * restrict retval, char * restrict dst, rsize_t dstmax, const wchar_t * restrict src, rsize_t len);
errno_t memcpy_s(void * restrict s1, rsize_t s1max, const void * restrict s2, rsize_t n);
errno_t memmove_s(void *s1, rsize_t s1max, const void *s2, rsize_t n);
errno_t strcpy_s(char * restrict s1, rsize_t s1max, const char * restrict s2);
errno_t strncpy_s(char * restrict s1, rsize_t s1max, const char * restrict s2, rsize_t n);
errno_t strcat_s(char * restrict s1, rsize_t s1max, const char * restrict s2);
errno_t strncat_s(char * restrict s1, rsize_t s1max, const char * restrict s2, rsize_t n);
char *strtok_s(char * restrict s1, rsize_t * restrict s1max, const char * restrict s2, char ** restrict ptr);
int fwprintf_s(FILE * restrict stream, const wchar_t * restrict format, ...);
int fwscanf_s(FILE * restrict stream, const wchar_t * restrict format, ...);
int snwprintf_s(wchar_t * restrict s, rsize_t n, const wchar_t * restrict format, ...);
int swprintf_s(wchar_t * restrict s, rsize_t n, const wchar_t * restrict format, ...);
int swscanf_s(const wchar_t * restrict s, const wchar_t * restrict format, ...);
int vfwprintf_s(FILE * restrict stream, const wchar_t * restrict format, va_list arg);
int vfwscanf_s(FILE * restrict stream, const wchar_t * restrict format, va_list arg);
int vsnwprintf_s(wchar_t * restrict s, rsize_t n, const wchar_t * restrict format, va_list arg);
int vswprintf_s(wchar_t * restrict s, rsize_t n, const wchar_t * restrict format, va_list arg);
int vswscanf_s(const wchar_t * restrict s, const wchar_t * restrict format, va_list arg);
int vwprintf_s(const wchar_t * restrict format, va_list arg);
int vwscanf_s(const wchar_t * restrict format, va_list arg);
int wprintf_s(const wchar_t * restrict format, ...);
int wscanf_s(const wchar_t * restrict format, ...);
\end{lstlisting}

\begin{lstlisting}[
    label={list:not-impl-funcs},
    caption={Not Implemented Safer C Library API},%
    style=patchStyle
    ]
char *gets_s(char *s, rsize_t n);
errno_t getenv_s(size_t * restrict len, char * restrict value, rsize_t maxsize, const char * restrict name);
void *bsearch_s(const void *key, const void *base, rsize_t nmemb, rsize_t size, int (*compar)(const void *k, const void *y, void *context), void *context);
errno_t qsort_s(void *base, rsize_t nmemb, rsize_t size, int (*compar)(const void *x, const void *y, void *context), void *context);
errno_t strerror_s(char *s, rsize_t maxsize, errno_t errnum);
size_t strerrorlen_s(errno_t errnum);
size_t strnlen_s(const char *s, size_t maxsize);
errno_t asctime_s(char *s, rsize_t maxsize, const struct tm *timeptr);
errno_t ctime_s(char *s, rsize_t maxsize, const time_t *timer);
struct tm *gmtime_s(const time_t * restrict timer, struct tm * restrict result);
struct tm *localtime_s(const time_t * restrict timer, struct tm * restrict result);
errno_t wcscpy_s(wchar_t * restrict s1, rsize_t s1max, const wchar_t * restrict s2);
errno_t wcsncpy_s(wchar_t * restrict s1, rsize_t s1max, const wchar_t * restrict s2, rsize_t n);
errno_t wmemcpy_s(wchar_t * restrict s1, rsize_t s1max, const wchar_t * restrict s2, rsize_t n);
errno_t wmemmove_s(wchar_t *s1, rsize_t s1max, const wchar_t *s2, rsize_t n);
errno_t wcscat_s(wchar_t * restrict s1, rsize_t s1max, const wchar_t * restrict s2);
errno_t wcsncat_s(wchar_t * restrict s1, rsize_t s1max, const wchar_t * restrict s2, rsize_t n);
wchar_t *wcstok_s(wchar_t * restrict s1, rsize_t * restrict s1max, const wchar_t * restrict s2, wchar_t ** restrict ptr);
size_t wcsnlen_s(const wchar_t *s, size_t maxsize);
errno_t wcrtomb_s(size_t * restrict retval, char * restrict s, rsize_t smax, wchar_t wc, mbstate_t * restrict ps);
errno_t mbsrtowcs_s(size_t * restrict retval, wchar_t * restrict dst, rsize_t dstmax, const char ** restrict src, rsize_t len, mbstate_t * restrict ps);
errno_t wcsrtombs_s(size_t * restrict retval, char * restrict dst, rsize_t dstmax, const wchar_t **restrict src, rsize_t len, mbstate_t * restrict ps);
\end{lstlisting}

\begin{lstlisting}[
    label={list:partial-impl-funcs},
    caption={Partially Implemented Safer C Library API},%
    style=patchStyle
    ]
errno_t tmpfile_s(FILE * restrict * restrict streamptr);
errno_t tmpnam_s(char *s, rsize_t maxsize);
errno_t fopen_s(FILE * restrict * restrict streamptr, const char * restrict filename, const char * restrict mode);
errno_t freopen_s(FILE * restrict * restrict newstreamptr, const char * restrict filename, const char * restrict mode, FILE * restrict stream);
\end{lstlisting}

\subsection{Private Constraint Handling API}
\index{Implemented API!Private Constraint Handling}

\subsubsection{Data Types}

Added the following data types:
\api{param\_validation\_status\_t},
\api{param\_range\_t},
\api{object\_range\_t} that were necessary to add.

\subsubsection{Functions}

The functions in \xl{list:our-impl-funcs} were to the large extend implemented
by our team as of this writing. Likewise, \xl{list:our-not-impl-funcs} lists
API not yet addressed.

\begin{lstlisting}[
    label={list:our-impl-funcs},
    caption={Implemented Constraint Handling API},%
    style=patchStyle
    ]
errno_t __validate_n_format(const char * restrict format, va_list args);
errno_t __validate_s_format(const char * restrict format, va_list args);
errno_t __validate_sn_format(const char * restrict format, va_list arg);


/* constraint validator; constraint_validator_s.c */

errno_t __error_out(bool errflag, param_validation_status_t * restrict status);
errno_t __constraint_validator_s(param_validation_status_t * restrict status);

errno_t
__constraint_validator_object_overlap(const char * restrict function_name, const char * restrict parameterNames, const void * restrict object1Start, const size_t object1Size, const void * restrict object2Start, const size_t object2Size);

errno_t
__constraint_validator_value_inrange(const char * restrict function_name, const char * restrict parameterName, const size_t value, const size_t lowerBound, const size_t upperBound);

errno_t
__constraint_validator_not_null(const char * restrict function_name, const char * restrict parameter_name, const void * restrict value_ptr);

errno_t
__constraint_validator_s_format(const char * restrict function_name, const char * restrict parameter_name, const char * restrict format, const va_list args);

errno_t
__constraint_validator_n_format(const char * restrict function_name, const char * restrict parameter_name, const char * restrict format, const va_list args);

errno_t
__constraint_validator_not_zero(const char * restrict function_name, const char * restrict parameter_name, const size_t value);

errno_t
__constraint_validator_rsize_limit(const char * restrict function_name, const char * restrict parameter_name, const rsize_t value);

void
__report_constraint_violation_end_of_token_not_present(const char * restrict function_name, const char * restrict parameter_name);
\end{lstlisting}

\begin{lstlisting}[
    label={list:our-not-impl-funcs},
    caption={Not Implemented Constraint Handling API},%
    style=patchStyle
    ]
errno_t __validate_args_not_null(const char * restrict format, va_list args);

errno_t
__constraint_validator_not_null_args(const char * restrict function_name, const char * restrict parameter_name, const char * restrict format, const va_list args);
\end{lstlisting}

\section{Constraint Handling In Action -- stdio}
\label{sect:example-2}

Test code is in \xl{list:stdio-test}.

\begin{lstlisting}[
    label={list:stdio-test},
    caption={Example of a Test Program for stdio to test and reject invalid \%s and \%n cases.},%
    style=patchStyle
    ]
/*
 * Sloppy Programming Test Cases
 */

#define __STDC_WANT_LIB_EXT1__ 1
#include "stdio.h"
#include "stdlib.h"

#include <unistd.h>
#include <string.h>

int
main(int argc, char** argv)
{
	int valid = 1;

	set_constraint_handler_s(ignore_handler_s);

	if(argc > 1)
	{
		printf("Sloppy programming zone: [[%s]]\n", argv[1]);
		printf_s(argv[1]);
		printf("\n\n");
	}

	printf_s("valid s = [%s]\n", "valid");
	printf_s("  valid n1 = [%%%%n]\n\n", &valid);
	printf_s("invalid n2 = [%%%n]\n\n", &valid);
	printf_s("  valid n3 = [%%n]\n\n", &valid);
	printf_s("invalid n4 = [%n]\n\n", &valid);

	printf_s("invalid s = [%s]\n", NULL);
	printf_s("invalid n = [%n]\n\n", &argc);

	printf("return value for %%n: [%d]\n", printf_s("%n\n", &argc));
	printf("return value for %%s: [%d]\n", printf_s("%s\n", NULL));

	return 0;
}

/* EOF */
\end{lstlisting}

\clearpage
Output is in \xl{list:stdio-out}.

\begin{lstlisting}[
    label={list:stdio-out},
    caption={Output},%
    style=patchStyle
    ]
bash-2.05b$ test/stdio/test %n
Sloppy programming zone: [[%n]]
printf_s(): invalid format parameter (%n is disallowed) : format/args %n


valid s = [%s]
valid s = [valid]
  valid n1 = [n]

  valid n1 = [%%n]

invalid n2 = [printf_s(): invalid format parameter (%n is disallowed) : format/args %n
  valid n3 = [n]

  valid n3 = [%n]

invalid n4 = [printf_s(): invalid format parameter (%n is disallowed) : format/args %n
printf_s(): invalid format parameter (NULL argument for %s) : format/args null %s
invalid n = [printf_s(): invalid format parameter (%n is disallowed) : format/args %n
printf_s(): invalid format parameter (%n is disallowed) : format/args %n
return value for %n: [22]
printf_s(): invalid format parameter (NULL argument for %s) : format/args null %s
return value for %s: [22]
bash-2.05b$
\end{lstlisting}

\section{Constraint Handling In Action -- string}
\label{sect:example-2}

Test code is in \xl{list:string-test}.

\begin{lstlisting}[
    label={list:string-test},
    caption={Example of a Test Program for string to test and reject invalid cases.},%
    style=patchStyle
    ]
#define __STDC_WANT_LIB_EXT1__ 1
#include "string.h"
#include <errno.h>
#include <stdio.h>
#include <string.h>
#include "stdlib.h"

int main(int argc, char** argv){
	char buffer1[1024];
	char buffer2[1024];
	
	set_constraint_handler_s(ignore_handler_s);
	
	/*Failure tests*/
	printf("strcpy_s failure test:\t");
	strcpy_s(buffer1, 1024, NULL);
	printf("strncpy_s failure test:\t");
	strncpy_s(buffer1, 10, buffer2, 50);
	printf("strncat_s failure test:\t");
	strcat_s(buffer1, -1, buffer2);
	printf("strncat_s failure test:\t");
	strncat_s(NULL, 1024, NULL, 50);
	printf("strncat_s failure test:\t");
	strncat_s(buffer1, 1024, buffer1-10, 50);
	printf("memcpy_s failure test:\t");
	memcpy_s(buffer1, 1024, buffer2, -1);
	printf("memmove_s failure test:\t");
	memmove_s(buffer1, 1023, NULL, 1024);

	/*Normal operation tests*/
	strcpy_s(buffer1, 1024, "test string");
	printf("strcpy_s: %s \n", buffer1);    
	strncpy_s(buffer2, 1024, buffer1, 1024);
	printf("strncpy_s: %s \n", buffer2);
	strcat_s(buffer1, 1024, buffer2);
	printf("strcat_s: %s \n", buffer1);    
	...
	printf("strnlen_s(buffer1): %u \n", strlength);
	size_t errlen = strerrorlen_s(EINVAL);
	printf("strerrorlen_s(EINVAL): %u\n", errlen);
	strerror_s(buffer2, 1014, EINVAL);
	printf("%s\n", buffer2);
	rsize_t l= strnlen_s (buffer1, 100);
	char * strktokresult = NULL;
	char * token = strtok_s(buffer1, &l, " ", &strktokresult);
	printf("strtok_s token: %s, remaining length: %u, remaining substring: %s\n", token, l, strktokresult);
	token = strtok_s(NULL, &l, "gt", &strktokresult); /*Should be found*/
	...
	/*Move printfs to a real test file*/
	size_t s = strnlen_s("12345", 10);
	printf("size: %u\n", s);
}
\end{lstlisting}

\clearpage
Output is in \xl{list:string-out}.

\begin{lstlisting}[
    label={list:string-out},
    caption={Output},%
    style=patchStyle
    ]
bash-2.05b$ test/string/test
strcpy_s failure test:  strcpy_s(): has invalid NULL pointer argument : s2
strncpy_s failure test: strncat_s failure test: strcat_s(): rsize_t value exceeds RSIZE_MAX : s1max
strncat_s failure test: strncat_s(): has invalid NULL pointer argument : s1
strncat_s failure test: strncat_s(): two data structures overlap in memory : s1 and s2
memcpy_s failure test:  memcpy_s(): rsize_t value exceeds RSIZE_MAX : n
memmove_s failure test: memmove_s(): has invalid NULL pointer argument : s2
strcpy_s: test string
strncpy_s: test string
strcat_s: test stringtest string
strncat_s: test stringtest stringtest string
memmove_s: test stringtest stringtest string
memcpy_s: test stringtest stringtest string
strnlen_s(buffer1): 33
strerrorlen_s(EINVAL): 0
test string
strtok_s token: test, remaining length: 28, remaining substring: stringtest stringtest string
strtok_s token: strin, remaining length: 21, remaining substring: est stringtest string
strtok_s(): token end not found within defined bounds : *ptr
strtok_s token: est stringtest string, remaining length: 0, remaining substring:
size: 5
bash-2.05b$
\end{lstlisting}


	\chapter{Conclusions}
\index{Conclusions}

Here were briefly address the following topics:

\begin{itemize}
\item Difficulties
\item Limitations
\item Acknowledgments
\item Future Work
\end{itemize}

\section{Summary of the Difficulties}

\begin{enumerate}
\item Parsing/processing of varargs and \%n in particular
\item Deciding on default values
\item Implementing \api{strtok\_s} and its wide character equivalent
\end{enumerate}

\section{Limitations So Far}

\begin{itemize}
\item Incomplete implementation (of approx. 45\%) of the entire API
\item Lack of thorough testing for all the implemented API
\end{itemize}

\section{Acknowledgments}

\begin{itemize}
\item Dr. Prabir Bhattacharya
\item Dr. Mourad Debbabi
\item ISO
\item Open Source Community and the GLIBC Team \cite{glibc}
\end{itemize}

\section{Future Work}

This project, being a derivative of the standard C library, will see a major effort put into the testing of
the library. Furthermore, a large part of the project (transforming the obsolete calls to wrappers to the
newer ones) makes the assumption that the buffer free size is easily obtainable, an assumption which
may not necessarily hold true in all circumstances. As such, it is possible that we will have to develop
novel algorithms to ensure that this portability is possible, or it may also be that this portability is not
100\% attainable.
Furthermore, we will need to investigate good potential software for performance and security testing
of our improved solution. Thus, we will focus on:

\begin{itemize}
\item Completion of implementation,
\item Addition more comprehensive test cases by the developers and the OSS community,
\item Application for EAL5,
\item Inclusion into the Linux kernel as a standard.
\end{itemize}


	\addcontentsline{toc}{chapter}{Bibliography}

\bibliography{report}
\bibliographystyle{alpha}



	\printindex
\end{document}